\documentclass[a4paper, twocolumn,superscriptaddress,prd,showpacs,preprintnumbers]{revtex4}
\usepackage{epsf} 
\usepackage{amsmath,color}
\usepackage{amssymb}
\usepackage{epsfig}
\usepackage{latexsym}
\usepackage{amsfonts}
\usepackage{dcolumn}
\usepackage{varioref}
\usepackage{ifthen}
\usepackage{graphicx}
\usepackage{amssymb,amsmath,amsthm}
\begin{document}
\parindent 0mm 
\setlength{\parskip}{\baselineskip} 
\thispagestyle{empty}
\pagenumbering{arabic} 
\setcounter{page}{1}
\mbox{ }
\preprint{UCT-TP-300/14}
\newline
\newline
\title{Weinberg sum rules at finite temperature}
\author{Alejandro Ayala}
\affiliation{Instituto de Ciencias Nucleares, Universidad Nacional
Aut\'onoma de M\'exico, Apartado Postal 70-543, M\'exico Distrito Federal
04510, Mexico}
\affiliation{Centre for Theoretical \& Mathematical Physics and Department of Physics, University of
Cape Town, Rondebosch 7700, South Africa}
\author{C. A. Dominguez}
\affiliation{Centre for Theoretical \& Mathematical Physics and Department of Physics, University of
Cape Town, Rondebosch 7700, South Africa}
\author{M. Loewe}
\affiliation{Centre for Theoretical \& Mathematical Physics and Department of Physics, University of
Cape Town, Rondebosch 7700, South Africa}
\affiliation{Instituto de F\'{i}sica, Pontificia Universidad Cat\'{o}lica de Chile, Casilla 306, Santiago 22, Chile}
\author{Y. Zhang}
\affiliation{CSIR National Laser Centre, PO Box 395, Pretoria 0001, South Africa}

\date{\today}
\begin{abstract}
The saturation of the  two Weinberg sum rules is studied at finite temperature, using recent independent QCD sum rule results for the thermal behaviour of
hadronic parameters in the vector and axial-vector channels. Both sum rules are very well satisfied from $T=0$ up to $T/T_c \simeq 0.7-0.8$. At higher temperatures close to $T_c$
a hadronic, pion-loop contribution in the space-like region proportional to $T^2$, present at leading order in the vector but not in the axial-vector channel, induces an asymmetry leading to a small deviation. In this region, though, QCD sum rules for the hadronic parameters begin to have no solutions, as the hadronic widths of the $\rho$ and the $a_1$ mesons diverge signaling deconfinement. Close to, and at $T=T_c$ there are no pions left in the medium and chiral symmetry is restored, so that the sum rules are trivially satisfied.
\end{abstract}
\maketitle
%

\section{Introduction}
The two Weinberg sum rules (WSR) (at T=0) \cite{WSR} were first derived in the framework of chiral $SU(2) \times SU(2)$ symmetry and current algebra, and read
\begin{equation}
W_1 \equiv 
\int\limits_{0}^{\infty} ds \,\frac{1}{\pi}\, [{\mbox{Im}} \Pi_V(s) - {\mbox{Im}} \Pi_A(s) ] = 2 \, f_\pi^2\;,
\end{equation}
\begin{equation}
W_2 \equiv 
\int\limits_{0}^{\infty} ds \,s \,\frac{1}{\pi}\, [{\mbox{Im}} \Pi_V(s) - {\mbox{Im}} \Pi_A(s) ] =0 \;,
\end{equation}
where $f_\pi = 92.21 \pm 0.14 \, {\mbox{{MeV}}}$ \cite{PDG}, and 
\begin{eqnarray}
\Pi_{\mu\nu}^{VV} (q^{2})   &=& i \, \int\; d^{4} \, x \, e^{i q x} \,
<0|T( V_{\mu}(x)   V_{\nu}^{\dagger}(0))|0> \nonumber \\ [.3cm]
&=& (-g_{\mu\nu}\, q^2 + q_\mu q_\nu) \, \Pi_V(q^2)  \; ,
\end{eqnarray}
\begin{eqnarray}
\Pi_{\mu\nu}^{AA} (q^{2})   &=& i \, \int\; d^{4} \, x \, e^{i q x} \,
<0|T( A_{\mu}(x)   A_{\nu}^{\dagger}(0))|0> \nonumber \\ [.3cm]
&=& -g_{\mu\nu}\;  \Pi_1(q^2) - q_\mu q_\nu\; \Pi_A(q^2)\,,
\end{eqnarray}
with $V_\mu(x) = :\bar{d}(x) \gamma_\mu \, u(x):$ the conserved vector current in the chiral limit, $A_\mu(x) = :\bar{d}(x) \gamma_\mu \gamma_5 \, u(x):$ the axial-vector current, and  $q_\mu = (\omega, \vec{q})$ the four-momentum carried by the currents. The functions $\Pi_{V,A}(q^2)$ are free of kinematical singularities, thus satisfying dispersion relations, and in perturbative QCD (PQCD) they are normalized as
\begin{equation}
	{\mbox{Im}} \,\Pi_{V}(q^2) = {\mbox{Im}} \,\Pi_A(q^2) = \frac{1}{4\pi}\left[ 1 + {\cal{O}} \left( \alpha_s(q^2)\right)\right]\,.
\end{equation}
In the framework of QCD Eqs. (1)-(2) become effectively finite energy sum rules (in the chiral limit: $m_u = m_d = 0$)
\begin{equation}
W_{n+1}(s_0) \equiv \int\limits_{0}^{s_0} ds \, s^n\, \frac{1}{\pi}\, [{\mbox{Im}} \Pi_V(s) - {\mbox{Im}} \Pi_A(s) ] = 2 \, f_\pi^2\, \delta_{n 0}\;,
\end{equation}

where $s_0 \simeq 1-3 \, {\mbox{GeV}^2}$ is the squared energy beyond which QCD is valid, and both sum rules have been combined. This result also follows from Cauchy's theorem in the complex squared energy plane, together with the assumption of quark-hadron duality, to wit. Given a Green function $\Pi(s)$, e.g. $\Pi_{V,A}(s)$, and an analytic function $f(s)$, Cauchy's theorem leads to
\begin{equation}
\int\limits_{0}^{s_0} ds \,f(s)\, \frac{1}{\pi}\, {\mbox{Im}} \Pi(s)= - \frac{1}{2 \pi i} \, \oint_{|s|=s_0} f(s)\, \Pi(s) \, ds \,,
\end{equation}
where the discontinuity across the real axis involves the hadronic sector, and the contour integral requires asymptotic information if the integration radius $s_0$ is large enough. According to quark-hadron (global) duality $\Pi(s)$ is assumed to be given by the QCD  operator product expansion (OPE). A warning about this assumption was first raised long ago in \cite{Shankar}, and is now referred to as potential duality violations associated with the QCD behaviour on or close to the positive real axis. Tests of the WSR were performed \cite{PS}-\cite{CAD-S} using experimental data on $\tau$-decays from the ARGUS collaboration at DESY \cite{ARGUS}, and later \cite{CAD-KS} from the ALEPH collaboration \cite{ALEPH1}. Both determinations led to a better saturation of the WSR if an integral kernel, e.g. a non-trivial $f(s)$ in Eq.(7), was introduced, e.g. such that $f(s_0)=0$. This was interpreted in \cite{CAD-KS} as a possible signal of duality violations, on account of a considerably improved saturation of the WSR obtained with a {\it pinched} kernel \cite{CAD-KS}, \cite{KM} $f(s) = (1 - s/s_0)$ , i.e.
\begin{eqnarray}
WP(s_0) &\equiv&
\int\limits_{0}^{s_0} ds \left( 1 - \frac{s}{s_0}\right) \frac{1}{\pi}\; [{\mbox{Im}} \Pi_V(s) - {\mbox{Im}} \Pi_A(s) ]\nonumber \\ [.3cm]
&=& 2\;  f_\pi^2 \,.
\end{eqnarray}

In this paper we discuss the extension of the WSR to finite temperature, which requires reliable  information on the thermal behaviour of the hadronic spectral functions. This information is available from results obtained from finite energy QCD sum rules (FESR) at finite $T$ in the vector \cite{AA} and independently in the axial-vector channel \cite{CAD}. 
\begin{figure}
\centering
\def\svgwidth{0.8\columnwidth}
\includegraphics[height=3.0in, width=3.5in]{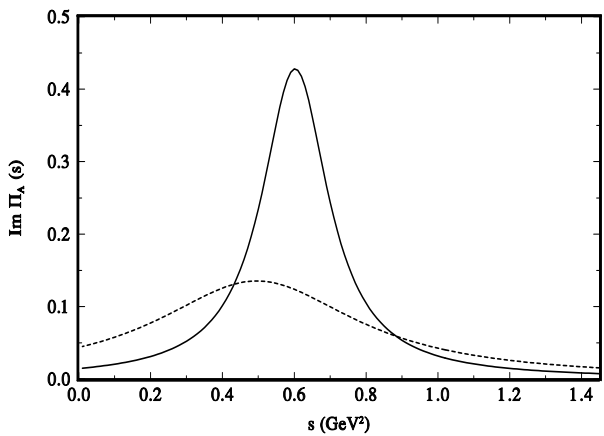}
\caption{{\protect\small{The vector spectral function at $T=0$, Eq. (9) (solid curve), and at $T= 175 \;{\mbox{MeV}}$ (dotted curve) with thermal parameters given in Eqs. (11)-(14).}}}
\end{figure}  
\begin{figure}
\centering
\def\svgwidth{0.8\columnwidth}
\includegraphics[height=3.0in, width=3.5in]{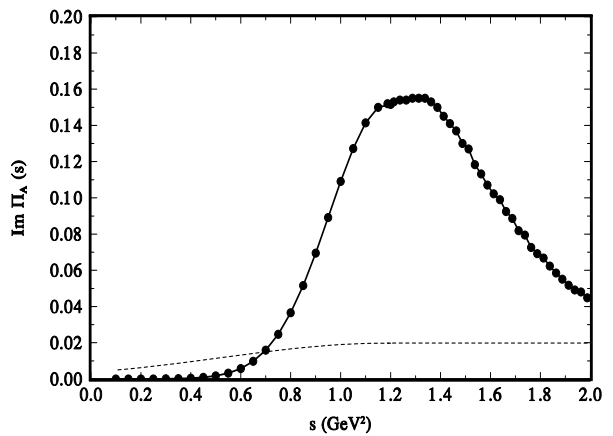}
\caption{{\protect\small{Solid curve is the axial-vector ($a_1$-resonance) spectral function at $T=0$ fitted to the ALEPH data \cite{ALEPH1}, shown with  error bars the size of the data points. Dotted curve is the spectral function at  $T= 175 \;{\mbox{MeV}}$ with thermal parameters given in Eqs. (16)-(18).}}}
\end{figure}  
 
\section{Hadronic spectral functions at finite temperature}
Starting with the hadronic vector spectral function, in the analysis of \cite{AA} it was parametrized by a Breit-Wigner form
\begin{equation}
	\frac{1}{\pi} {\mbox{Im}} \Pi_{V}^{(+)}(s,T) = \frac{1}{\pi}\, \frac{1}{f_\rho^2}\, \frac{M_\rho^3 \, \Gamma_\rho}{\left(s - M_\rho^2\right)^2 + M_\rho^2 \,\Gamma_\rho^2}\;, \label{BW}
\end{equation}
where the upper index (+) specifies the time-like region ($q^2 > 0$), and all parameters depend on the temperature, and are normalized at $T=0$ as  follows \cite{PDG}: $f_\rho(0)=5$, $M_\rho(0)=0.776$ GeV, and $\Gamma_\rho(0) = 0.145$ GeV. This parametrization was then  used in the first three FESR
\begin{eqnarray}
&(-)^{(N-1)}& C_{2N} \langle {\mathcal{\hat{O}}}_{2N}\rangle = 4 \pi^2 \int_0^{s_0} ds\, s^{N-1} \,\frac{1}{\pi} {\mbox{Im}} \Pi(s)|_{\mbox{\scriptsize
{HAD}}}
\nonumber \\ [.3cm]
&-& \frac{s_0^N}{N} \left[1+{\mathcal{O}}(\alpha_s)\right] \;\; (N=1,2,\cdots) \;,\label{FESR}
\end{eqnarray}
where the leading order vacuum condensates in the chiral limit are the dimension $d=4$ gluon condensate and the dimension $d=6$ four-quark condensate.
The FESR were then used  to determine $s_0 = 1.44 \;{\mbox{GeV}^2}$, i.e. the onset of PQCD, and to check that the resulting dimension $d=4$ and $d=6$ vacuum condensates were in agreement with independent determinations from data analyses. Alternative parametrizations were also considered but found to lead to similar results.
\begin{figure}
\centering
\def\svgwidth{0.8\columnwidth}
\includegraphics[height=3.0in, width=3.5in]{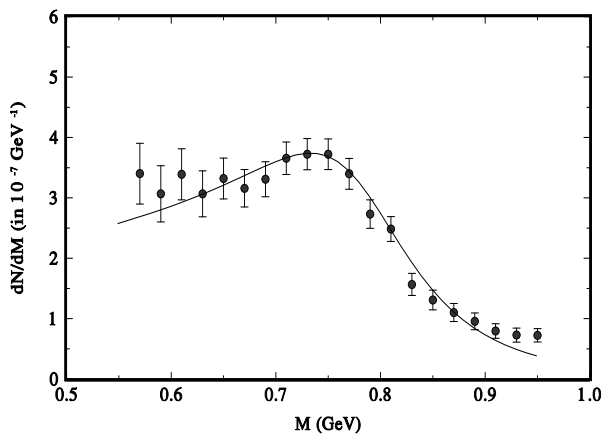}
\caption{{\protect\small{Experimental data on the dimuon production spectrum in In-In ion collisions around the $\rho$-peak from the NA60 Collaboration \cite{NA60}, compared with the prediction using the vector spectral function from \cite{AA} (solid line), reproduced here in Eqs.(9), (11)-(14).}}}
\end{figure}  

Extending this analysis to finite temperature gave the following results
\begin{equation}
\frac{s_0(T)}{s_0(0)} = 1.0 - 0.5667 \left(\frac{T}{T_c}\right)^{a} - 4.347
\left(\frac{T}{T_c}\right)^{b}\,,
\end{equation}
with $a=11.38$, $b=68.41$, $T_c = 197 \, \mbox{{MeV}}$ and
\begin{equation}
\frac{\Gamma_\rho(T)}{\Gamma_\rho(0)} = \frac{1}{\left[1 - \left(\frac{T}{T_c}\right)^3 \right]}\;,
\end{equation}
\begin{equation}
\frac{M_\rho(T)}{M_\rho(0)} = 1 - \left(\frac{T}{T^*_M}\right)^{10}  \;,
\end{equation}
where $T^*_M = 222\; {\mbox{MeV}}$,  allowed to vary in the  range $T^*_M = 210 - 240\; {\mbox{MeV}}$, and finally
\begin{equation}
\frac{f_\rho(T)}{f_\rho(0)} = 1.0 - 0.3901 \left(\frac{T}{T_c}\right)^{c} + 0.04155
\left(\frac{T}{T_c}\right)^{d}\,,
\end{equation}
with $c= 10.75$ and $d=1.269$. \\

Turning to the axial-vector hadronic resonance parametrization, a fit was made to the ALEPH data on $\tau$-decay in order to use as the $T=0$ normalization. As in the vector channel, the first three FESR were employed to determine $s_0$ and to check consistency of the results for the dimension $d=4,6$ condensates. The parametrization leading to the best $\chi^2$ is given by \cite{AA}
\begin{eqnarray}
\frac{1}{\pi} {\mbox{Im}} \Pi_A(s)|_{a_1} &=& C\, f_{a_1} \exp\left[-\left(\frac{s- M_{a_1}^2}{\Gamma^2_{a_1}}\right)^2\right] 
\nonumber \\ [.3cm]
&(0& \leq s \leq 1.2\, {\mbox{GeV}}^2)\;,
\end{eqnarray}
\begin{eqnarray}
\frac{1}{\pi} {\mbox{Im}} \Pi_A(s)|_{a_1} &=& C\, f_{a_1} \exp\left[-\left(\frac{1.2\; {\mbox{GeV}}^2 - M_{a_1}^2}{\Gamma^2_{a_1}}\right)^2\right] 
\nonumber \\ [.3cm]
&&(1.2\, {\mbox{GeV}}^2 \leq s \leq  1.45\,{\mbox{GeV}}^2)\;,
\end{eqnarray}
\begin{eqnarray}
\frac{1}{\pi} {\mbox{Im}} \Pi_A(s)|_{a_1} &=& C\, f_{a_1} \exp\left[-\left(\frac{s- M_{a_1}^2}{\Gamma^2_{a_1}}\right)^2\right] 
\nonumber \\ [.3cm]
&&(1.45\, {\mbox{GeV}}^2 \leq s \leq M_\tau^2)\;,
\end{eqnarray}
where $M_{a_1} = 1.0891 \; {\mbox{GeV}}$, $\Gamma_{a_1} = 568.78\; {\mbox{GeV}}$, $C= 0.662$ and $f_{a_1}= 0.073$ (the latter two parameters were split to facilitate comparison between $f_{a_1}$ and $f_{\rho}$, for readers used to zero-width resonance saturation of the WSR). Regarding Eqs.(15)-(17), it is well known that this spectral function cannot be fitted with a normal Breit-Wigner expression, mostly due to the large width of the $a_1$, and some background contamination. For instance, the early ARGUS data \cite{ARGUS} was fitted with a Breit-Wigner modulated by a high degree polynomial \cite{CADSOLA}.
Solving the FESR the onset of PQCD turns out to be identical to that in the vector channel, i.e. $s_0 = 1.44 \;{\mbox{GeV}^2}$, so that Eq.(17) is never needed. The finite temperature results for $s_0(T)$, $f_\pi(T)$, $f_{a_1}(T)$, and $\Gamma_{a_1}(T)$ can be written generically as
\begin{equation}
\frac{Y(T)}{Y(0)} \,=\, 1 \, + \, a_1\, \left(\frac{T}{T_c}\right)^{b_1}\, + \,a_2 \, \left(\frac{T}{T_c}\right)^{b_1} \;, 
\end{equation}
where the various coefficients are given in Table I. The $a_1$ mass hardly changes with temperature, so that it was kept constant. This behaviour can be ascribed to the very large width of the $a_1$ resonance. It should be noticed the (different) way the leptonic decay constants enter their respective spectral functions. However, the full $T$-dependence of these functions is the result of the aggregate behaviour of all their parameters. This $T$-dependence is consistent with a quark deconfinement scenario in both channels.\\

The parametrization of the spectral function in the vector channel was later used to predict the dimuon production spectrum in heavy ion collisions at high energies \cite{Dimuon}. Excellent agreement was obtained after comparing with data on In-In collision around the $\rho$-meson peak from the NA60 Collaboration \cite{NA60}. Figure 3 shows the experimental data and the prediction using the vector spectral function from \cite{AA}, as reproduced here in Eqs.(9), (11)-(14).\\

It should be clear from Figs. (1) and (2) that the vector spectral function evolves with increasing $T$ independently of the axial-vector one. The latter involves, in addition to the $a_1$ resonance, the pion pole given by $f_\pi^2(T) \propto |\langle \bar{q} \,q\rangle|$. This contribution vanishes at $T=T_c$, signaling chiral-symmetry restoration, i.e. a transition from a Nambu-Goldstone to a Wigner-Weyl realization of chiral $SU(2) \times SU(2)$ symmetry. It should be recalled that the thermal quark condensate from lattice QCD has been used as input to the FESR in the axial-vector channel in \cite{CAD}.
As shown in \cite{CAD} the $T$-dependence of  the ratio$f_\pi(T)^2/f_\pi(0)^2$ follows closely that of $s_0(T)/s_0(0)$, vanishing at the same temperature within the accuracy of the method. This independent and clearly very different approach to $T=T_c$ exhibited by the thermal spectral functions does not suggest any potential chiral mixing.
\squeezetable
\begin{table}
\begin{ruledtabular}
\begin{tabular}{ccccc}
\multicolumn{4}{r}
{Coefficients in Eq. (18)} \\
\cline{2-5}
\noalign{\smallskip}
 Parameter  & $a_1$ &$a_2$  & $b_1$ &$b_2$   \\
\hline
\noalign{\smallskip}
$s_0(T)$ & - 28.5 & -0.6689 & 35.60 & 3.93  \\
$f_\pi(T)$ &- 0.2924 & - 0.7557 & 73.43 & 11.08   \\
$f_{a_1}(T)$ & - 19.34 & 14.27 & 7.716 & 6.153 \\
$\Gamma_{a_1}(T)$&2.323 & 1.207 & 20.24 & 7.869\\
\end{tabular}
\caption{\footnotesize{The values of the coefficients entering Eq. (18).}}
\end{ruledtabular}
\end{table}
 
\begin{figure}
\centering
\def\svgwidth{0.8\columnwidth}
\includegraphics[height=3.0in, width=3.5in]{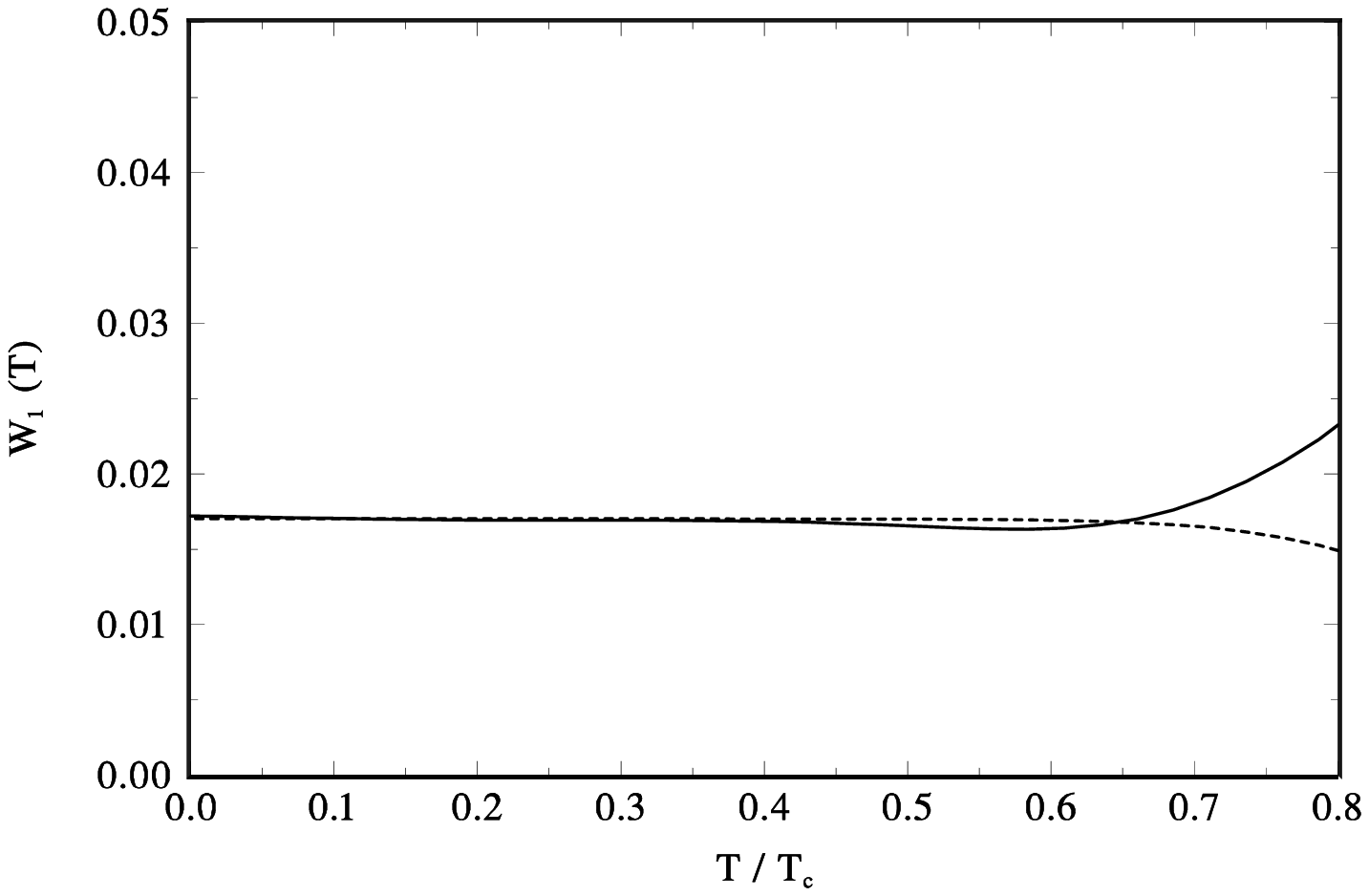}
\caption{{\protect\small{The first WSR, Eq.(6),  at finite T. Solid (dash) line is the left (right) hand side of Eq.(6). The divergence at high $T$ is caused by the asymmetric hadronic {\it{scattering}} contribution in the space-like region ($q^2 <0$), which disappears at deconfinement ($T=T_c$).}}} 
\end{figure}

\begin{figure}
\centering
\includegraphics[height=3.0in, width=3.5in]{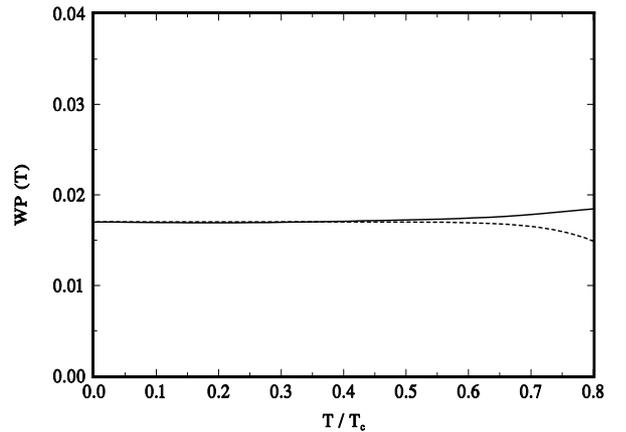}
\caption{{\protect\small{The pinched WSR, Eq. (8), at finite $T$. Solid (dash) line is the left (right) hand side of Eq.(6). The divergence at high $T$ is caused by the asymmetric hadronic {\it{scattering}} contribution in the space-like region ($q^2 <0$), which disappears at deconfinement ($T=T_c$).}}}
\end{figure}
In the QCD sector the PQCD spectral functions for both $q^2 > 0$ ({\it {annihilation}}) and $q^2 < 0$ ({\it {scattering}}), as well as the dimension $d=4$ gluon condensate are chiral symmetric, thus not contributing to Eq.(6). In the hadronic sector for $q^2 < 0$ ({\it {scattering}}) there is an asymmetric contribution from pion loops, which in the vector channel at leading order is \cite{BS}
\begin{eqnarray}
	\frac{1}{\pi}{\mbox{Im}}\, \Pi^{(-)}_V|_{HAD}(s,T) &=& \frac{2}{3\pi^2} \;\delta(s)\int_0^\infty y\, n_B\left(\frac{y}{T}\right)dy \nonumber \\ [.3cm]
	&=& \frac{T^2}{9} \; \delta(s) \;,
\end{eqnarray}
where $n_B(z) = 1/(e^{z}-1)$ is the Bose thermal function. The corresponding contribution in the axial-vector channel is a higher order two-loop effect, which is traditionally  neglected. 

\section{Results and conclusions}

Starting at $T=0$ and using the results from the three FESR obtained in \cite{AA}-\cite{CAD}, i.e. $s_0(0)|_V = s_0(0)|_A = 1.44\;{\mbox{GeV}^2} $ and the parametrizations Eq.(9) and Eqs.(15)-(17), the WSR, Eq.(6), and the pinched WSR, Eq.(8),  are satisfied to better than 1\%. The corresponding results at finite $T$ are shown in Fig. 4 for the first WSR, $W_1(T)$,  Eq.(6), and in Fig. 5 for the pinched WSR, $WP(T)$, Eq. (8). The solid (dash) lines correspond to the left (right) hand sides of the sum rules. Clearly, the upper limit of integration in the WSR, i.e. $s_0(T)$, is not strictly chiral-symmetric, as seen from Eqs.(11) and (18).
The divergence between the left and the right hand sides of the WSR at high $T$ is due to the asymmetric hadronic {\it{scattering}} term present in the vector spectral function, but loop suppressed in the axial-vector channel. The importance of this term is short lived, as it must disappear at $T=T_c$, since there would be no longer pions in the medium. In any case, since the source of this contribution is fully understood, it should not be interpreted as an invalidation of the thermal WSR.\\

 As indicated in Figs. 4-5 the WSR are very well satisfied up to $T/T_c \simeq 0.75$. The source of the departure between the left hand side of the WSR and its right hand side, i.e. $f_\pi^2(T)$, or alternatively the thermal quark condensate $|\langle \bar{q} q\rangle(T)|$, is well understood. It is due to the two-pion loop contribution to the vector spectral function, growing like $T^2$, but disappearing at $T=T_c$. If this contribution would have been chiral symmetric, then it would have canceled out and the WSR would have been fully satisfied up to $T=T_c$.\\ 
 
The results for the  behaviour of the WSR in the full temperature range can be interpreted as follows. First of all, at $T=0$ the WSR make the highly non-trivial statement that the integrated difference between the vector and the axial-vector (including the pion) spectral functions vanishes. These functions could not be more different, i.e. the vector correlator has no pole, and the ground state resonance is narrow, while the axial-vector involves a pole and a very broad resonance almost twice as heavy. Next, there is no a-priori reason for the WSR to be satisfied at finite temperature.  Their original derivation was done in the context of current algebra and chiral $SU(2) \times SU(2)$ symmetry. Today we reinterpret them in the framework of QCD, which does exhibit this chiral symmetry in the limit of vanishing light quark masses. In principle, an extension to finite $T$ could lead to more than one scenario, including one in which the sum rules are not satisfied at all, except at $T=T_c$ where they must be trivially satisfied. The results obtained from thermal QCD FESR \cite{AA}-\cite{CAD} clearly indicate that the vector spectral function evolves with $T$ independently of the axial-vector spectral function. While both disappear at $T=T_c$, they do so following their own distinctive thermal evolution. In other words, there is no a-priori relation between these two channels, unless one were to demand it. Given the successful saturation of the WSR there seems to be no reason to justify such a demand. The behaviour of the spectral function parameters points to deconfinement, as the onset of PQCD, $s_0(T)$, and the resonance couplings decrease, while the widths increase substantially with increasing $T$. This results in the $\rho$ and $a_1$ resonances  becoming progressively broader, with decreasing couplings to their respective currents. At the same time, the pion decay constant/quark condensate decreases with increasing $T$, vanishing at $T_c$. The thermal behaviour of the hadron mass is irrelevant in the context of the approach to deconfinement. In fact, the mass only specifies the position of the real part of the resonance Green function in the complex squared energy $s$-plane. Its moving upwards or downwards with temperature is no indication of a deconfinement phase transition. It is the imaginary part of the Green function, i.e. the hadronic width, the relevant parameter in this context. Its considerable increase with increasing temperature thus serves as a phenomenological signal for deconfinement. \\

Alternative analyses of the thermal WSR have been performed recently \cite{recent}, with somewhat different emphasis, e.g. in connection with potential chiral mixing.

\section{Acknowledgments} 
This work has been supported in part by DGAPA-UNAM under grant PAPIIT-IN103811, CONACyT-Mexico under Grant No. 128534, the National Research Foundation (South Africa), the University of Cape Town Research Committee, and FONDECyT (Chile) under Grants No. 1130056 and No. 1120770.

\end{document}